\documentclass[12pt]{article}

\usepackage[english]{babel}
\usepackage{graphicx,psfrag,epsf}
    \usepackage[T1]{fontenc}
\usepackage{textcomp}
\usepackage{amsmath}
\usepackage{amssymb}
\usepackage{amsthm}
\usepackage{xcolor}
\usepackage{hyperref}
\usepackage{subcaption}
\usepackage{natbib}
\usepackage{booktabs}
\usepackage{longtable}
\usepackage{tikz}
\usetikzlibrary{external} 
\usepackage{enumitem}
\usepackage[margin=1in]{geometry}

\usepackage{setspace}

\defcitealias{Schorfheide2015}{SS15}


\usepackage[symbol]{footmisc}

\usepackage{environ}
\makeatletter
\newsavebox{\measure@tikzpicture}
\NewEnviron{scaletikzpicturetowidth}[1]{%
  \def\tikz@width{#1}%
  \def\tikzscale{1}\begin{lrbox}{\measure@tikzpicture}%
  \BODY
  \end{lrbox}%
  \pgfmathparse{#1/\wd\measure@tikzpicture}%
  \edef\tikzscale{\pgfmathresult}%
  \BODY
}
\makeatother

\begin{document}

\begin{titlepage}
\renewcommand{\thefootnote}{\fnsymbol{footnote}}
\pagestyle{empty}
\vspace*{\fill}
  \vskip 1em%
  \begin{center}%
  \let \footnote \thanks
    {\LARGE 
\begin{spacing}{1.2}
Estimating Large Mixed-Frequency Bayesian VAR Models\footnotemark[1]
\end{spacing}}%
    \vskip 1.5em%
    {\large
      \lineskip .5em%
      \begin{tabular}[t]{c}%
        Sebastian Ankargren\footnotemark[2] and Paulina Jon\'{e}us\vspace{0.25cm}\\
        \emph{Department of Statistics, Uppsala University}
      \end{tabular}\par}%
    \vskip 1em%
  \end{center}%
  \par
  \vskip 1.5em

\begin{abstract}
We discuss the issue of estimating large-scale vector autoregressive (VAR) models with stochastic volatility in real-time situations where data are sampled at different frequencies. In the case of a large VAR with stochastic volatility, the mixed-frequency data warrant an additional step in the already computationally challenging Markov Chain Monte Carlo algorithm used to sample from the posterior distribution of the parameters. We suggest the use of a factor stochastic volatility model to capture a time-varying error covariance structure. Because the factor stochastic volatility model renders the equations of the VAR conditionally independent, settling for this particular stochastic volatility model comes with major computational benefits. First, we are able to improve upon the mixed-frequency simulation smoothing step by leveraging a univariate and adaptive filtering algorithm. Second, the regression parameters can be sampled equation-by-equation in parallel. These computational features of the model alleviate the computational burden and make it possible to move the mixed-frequency VAR to the high-dimensional regime. We illustrate the model by an application to US data using our mixed-frequency VAR with 20, 34 and 119 variables.
\end{abstract}

\footnotetext[1]{We are thankful for comments by Johan Lyhagen and seminar participants at the Department of Statistics, Uppsala University and Sveriges Riksbank. Part of this research was conducted while Ankargren was visiting Sveriges Riksbank, which is gratefully acknowledged. We also wish to thank the Swedish National Infrastructure for Computing (SNIC) through Uppsala Multidisciplinary Center for Advanced Computational Science (UPPMAX) under projects SNIC 2015/6-117 and 2018/8-361 for providing the necessary computational resources.}
\footnotetext[2]{Corresponding author: \mbox{S. Ankargren,~\href{mailto:sebastian.ankargren@statistics.uu.se}{sebastian.ankargren@statistics.uu.se}}}
\vspace*{\fill}

\end{titlepage}

\renewcommand{\thefootnote}{\arabic{footnote}}

\section{Introduction}
An almost unavoidable feature of macroeconomic datasets used in real-time forecasting is that data are inherently unbalanced. There are mainly two reasons for unbalancedness of datasets: different sampling frequencies, and publication delays causing missing values at the end of the sample. For example, the gross domestic product (GDP), a key economic variable, is available quarterly, whereas many other central economic variables such as inflation, unemployment and industrial production are predominantly available monthly, usually with a delay of one or two months. Some variables, such as variables related to the stock market and interest rates, are often available daily, if not even more frequently. Typically, VAR models are estimated on a single-frequency basis and, consequently, the inclusion of a quarterly variable like GDP determines whether the model will be estimated on monthly or quarterly data.

Econometric models taking information in unequal frequencies into account, and thereby avoiding a loss of information stemming from aggregation to the lower frequency, have in recent years gained attention under the name of mixed-frequency methods. Multiple approaches are available, see for example \cite{Foroni2013} for a review. One of the methods put forward in the literature is the mixed-frequency vector autoregressive (VAR) model.

The use of factor models, as in the key contributions by \cite{Mariano2003,Mariano2010} is another way of handling mixed-frequency data that is particularly popular when the set of variables available is large; see also \cite{Marcellino2016}. Other proposed approaches include the Mixed data sampling (MIDAS) regression and MIDAS-VAR proposed by \cite{Ghysels2007} and \cite{Ghysels2016} respectively. MIDAS regressions cope with the issue of unequal frequencies by regressing a low-frequency dependent variable on its lags as well as on lags of high-frequency variables. In the observation-driven MIDAS VAR model, the vector of endogenous variables includes both high and low-frequency variables and are formulated in terms of observable data. In contrast to the present paper, they do not involve latent processes. In the same spirit, \cite{McCracken2015} proposed a high-dimensional variant of the MIDAS-VAR. They use a mixed-frequency Bayesian VAR estimated at the lowest common data frequency and find that mixed-frequency information is important for nowcasting.

Our focus is a situation in which GDP growth is the sole quarterly variable, with the remaining being monthly variables. What we investigate is how a large set of monthly variables can be used when modeling GDP growth in a mixed-frequency VAR. The method we employ in this paper is to cast the mixed-frequency model in a state-space form. By doing so, we can essentially interpolate the latent monthly values of the quarterly variables concurrently with making inference about the parameters in the model. \cite{Eraker2015} formulated a Bayesian VAR based on this idea and \cite{Schorfheide2015} proceeded with a Gibbs sampling approach based on simulation smoothing and forward-filtering, backward-smoothing along the lines of \cite{Carter1994}. \cite{Schorfheide2015} found that the model improved forecasting as compared to a quarterly VAR model. In an extension of \cite{Schorfheide2015}, \cite{Ankargren2018} developed a steady-state mixed-frequency VAR model for a real-time US dataset and arrived at the same conclusion. Evidence of improved forecasting performance was also presented by \cite{Gotz2018}.

In parallel with the development of mixed-frequency VARs, the empirical success of large VARs in forecasting macroeconomic variables has been highlighted in a number of papers, including among others \cite{Banbura2010} and \cite{Koop2013}. There are multiple options for working with VARs in a high-dimensional setting, where the number of parameters is large. Setting the level of shrinkage in relation the dimension of the model was discussed by \cite{Banbura2010}, who subsequently showed that larger models can improve forecasting ability when doing so. In more recent work, new methods drawing from the high-dimensional literature have been developed, see e.g. \cite{Korobilis2019} who developed a highly scalable estimation method for BVARs, \cite{Koop2019} who proposed a model based on ideas from the compressed regression literature and \cite{KastnerHuber2018,Follett2019,Huber2017} who used global-local shrinkage priors.

In addition to the usefulness of large VARs, seminal work by \cite{Cogley2005} and \cite{Primiceri2005} highlighted the importance of letting parameters and volatilities in VARs vary over time; see also \cite{Clark2011,DAgostino2013,Carriero2015,Carriero2016} for more forecasting-related studies. Different solutions have been presented to circumvent the computational difficulties regarding high dimensions in combination with time-varying volatilities. \cite{Carriero2019} extended earlier work on estimation of Bayesian VARs with asymmetric priors and time-varying volatilities to the high-dimensional setting, where the assumption of a specific common structure for the volatilities was relaxed and a simple triangularization of the VAR was used to reduce computational complexity. \cite{KastnerHuber2018} proposed an alternative route that instead used a factor stochastic volatility model that allows for co-movements in the error covariance structure when estimating large-dimensional VAR models.

In terms of mixed-frequency VARs with stochastic volatility, \cite{Cimadomo2016} extended earlier work on stochastic volatility to allow for data sampled at different frequencies. Due to quickly increasing computational complexity, their proposed model is restricted to include only a small number of variables and lags. \cite{Gotz2018}, also closely related to our work, proposed a model for a moderately large set of variables where the intercept and the common factor in the error variances are allowed to vary over time. Their application used 11 variables and 6 lags, and the authors stated that this model could be enlarged to up to 20 variables. In the case of more variables or when the length of the sample is increased, they mention a considerable obstacle in the running time. Using a standard normal inverse Wishart prior, \cite{Brave2019} demonstrated that a larger, 37-variable mixed-frequency VAR outperformed the smaller model of \cite{Schorfheide2015}.

Our contribution is that we add to this growing literature on large mixed-frequency VARs. We propose a large mixed-frequency VAR where we model the time-varying error covariance matrix by a factor stochastic volatility model along the lines of \cite{KastnerHuber2018}. In contrast to \cite{KastnerHuber2018}, we include the possibility to estimate the model on mixed-frequency data. Apart from being a parsimonious approach for modeling time-varying volatility, use of the factor stochastic volatility model comes with computational advantages that we exploit for estimating our model when dimensions are high. The factor stochastic volatility model renders the equations in the model conditionally independent. Consequently, the adaptive simulation smoother developed by \cite{Ankargren2019} can be improved further by coupling it with the univariate filtering methodology suggested by \cite{Koopman2000}. Moreover, the conditional independence of the equations means that we can sample the rows of the regression parameter matrices independently in parallel. These computational improvements ameliorate the computational efficiency of our Markov Chain Monte Carlo (MCMC) sampler, thereby enabling faster estimation for large-scale mixed-frequency VARs. The proposed model is illustrated using the FRED-MD database constructed by \cite{McCracken2016}, where we consider three models of increasing size with 20, 34 and 119 variables. In summary, the model and computational strategies we propose substantially reduce the computational burden induced by the model, and, as such, provide an efficient and feasible way of estimating VARs with mixed-frequency data and stochastic volatilities for models of dimensions that have up to now been missing in the literature. A benefit of our approach is that we do not assume a Kronecker structure for the prior variance of the regression parameters in contrast to e.g. \cite{Gotz2018,Ankargren2018}, who used the methodology presented by \cite{Carriero2016}. Consequently, the model that we will discuss permits asymmetric prior distributions and can therefore be used in conjunction with high-dimensional hierarchical shrinkage priors.

The rest of the paper is organized as follows. Section \ref{sec:background} introduces the mixed-frequency VAR, Section \ref{sec:application} presents our suggested approach for large mixed-frequency VARs, Section \ref{sec:uni} develops a fast simulation smoother exploiting the structure of the model, and Section \ref{sec:crisis} provides an empirical illustration. Section \ref{sec:conclusion} concludes.

\section{Mixed-Frequency VARs}
\label{sec:background}
The mixed-frequency VAR($p$) is specified at the monthly frequency as
\begin{align}
x_t=\Pi_c+\Pi_1x_{t-1}+\cdots+\Pi_px_{t-p}+u_t, \quad u_t \sim N(0, \Sigma_t) \label{hf}
\end{align}
where $x_t$, $\Pi_c$ and $u_t$ are $n\times 1$ and $\Pi_j$ are $n\times n$. The vector $x_t$ consists of both monthly and quarterly variables and is partitioned as $x_t=(x_{m,t}', x_{q,t}')'$, where the dimension of $x_{m, t}$ is $n_m\times 1$ and and $x_{q, t}$ is of dimension $n_q\times 1$, where $n=n_m+n_q$. 

In a mixed-frequency setup, the dataset that is needed for estimating the VAR is not fully observed. Instead, every month $t$ we observe $n_t\leq n$ variables collected in the vector $y_t=(y_{m, t}', y_{q,t}')'$, where $y_{m, t}$ and $y_{q,t}$ are the observed monthly and quarterly subsets of dimensions $n_{m, t}\times 1$ and $n_{q,t}\times 1$, respectively. The vector of observables, $y_t$, relates to the underlying $x_t$ through
\begin{equation}
\begin{aligned}
y_{m, t}&=S_{m, t} x_{m,t}\\
y_{q, t}&=S_{q, t}\Lambda_{qq} \begin{pmatrix}x_{q, t}\\ x_{q, t-1}\\ \vdots \\ x_{q, t-p+1}\end{pmatrix}
\end{aligned}\label{eq:obseq}
\end{equation}
where $S_{m,t}$ and $S_{q,t}$ are deterministic selection matrices of dimensions $n_{m, t}\times n_m$ and $n_{q, t}\times n_q$, respectively. The matrix $\Lambda_{qq}$ is an $n_q\times pn_q$ aggregation matrix that aggregates the latent monthly series into observed quarterly variables according to a selected aggregation scheme. We use data transformed to stationarity and therefore employ the triangular aggregation suggested by \cite{Mariano2003}, whence $y_{q, t}=\frac{1}{9}(x_{q, t}+2x_{q,t-1}+3x_{q,t-2}+2x_{q,t-3}+x_{q,t-4})$ when $t$ corresponds to an end-of-quarter month (with $y_{q,t}=\varnothing$ otherwise).

The posterior distribution that we are interested in is $p(\Pi, \Sigma, x|y)$.\footnote{We will throughout the paper make use of the notational convention that variables without subscripts refer to the whole set of that variable, i.e. $\Pi=\{\Pi_i : i = 1, \dots, p\}$.} \cite{Schorfheide2015} developed a Gibbs sampler for estimating the model using the normal-inverse Wishart prior distribution (see \citealp{Kadiyala1993,Kadiyala1997}) for the regression parameters and the constant error covariance matrix. The Gibbs sampler uses as one of its steps a simulation smoothing algorithm \citep{Carter1994,FruhwirthSchnatter1994} for making a draw from the conditional posterior distribution $p(X|Y, \Pi, \Sigma)$. The basic Gibbs sampler for the mixed-frequency VAR is to sample repeatedly from:
\begin{align}
&p(\Pi, \Sigma|x)\\
&p(X|\Pi, \Sigma, y).
\end{align}
The first step is standard in the literature conditional on $x$, and conditionally independent of $y$. Intuitively, the $p(x|\Pi, \Sigma, y)$ step imputes the underlying monthly series of all variables observed quarterly as well as monthly series with missing values at the end of the sample. Next, conditional on $x$, draws of $\Pi$ and $\Sigma$ can be produced as if $x$ were the real data, leading to standard procedures. Given the modular property of MCMC, it is therefore possible to extend the underlying VAR model using existing methods employed for fully-observed VARs by essentially adding the mixed-frequency sampling step to existing MCMC algorithms; see \cite{Cimadomo2016,Gotz2018,Ankargren2018} for examples. It is this modularity that we exploit in this paper.

\section{A Large Mixed-Frequency VAR with Factor Stochastic Volatility}
\label{sec:application}
In this paper, we estimate three mixed-frequency models of increasing size. We attempt to stay relatively close to models used in the previous literature and use pre-existing single-frequency models now augmented with mixed-frequency data. Our purpose for doing so is to model GDP growth using VARs and a large set of monthly data.

The three models that we use are based on models used by \cite{Koop2013} and \cite{Carriero2019}. \cite{Koop2013} used a medium-size quarterly VAR with 40 variables, all transformed to the quarterly frequency, and \cite{Carriero2019} used two monthly models with $n=20$ and $n=125$, respectively. After excluding discontinued and short series, we end up with three models of dimensions: $n=20$, $n=34$ and $n=119$, where each contains $n-1$ monthly series and quarterly GDP growth. For ease of presentation, the models are referred to as CCM-20, Koop-34 and CCM-119, respectively. The current section describes the data, some of their important characteristics, and some key aspects of the models. The more technical details, including the complete set of prior specifications, are relegated to Appendix \ref{app:prior}.
\label{sec:empirical}

\subsection{Data Structure and Publication Schemes}
The data we use are retrieved from FRED-MD \citep{McCracken2016}. We use the January 2019 vintage and collect a typical observational pattern of the included variables from the St. Louis Federal Reserve's database ALFRED. Because the Koop-34 model includes a number of series no longer published in the FRED-MD vintages, the Koop-34 model uses data through December 2014. The data are transformed according to the transformations suggested by \cite{McCracken2016} and thereafter standardized. See Appendix \ref{sec:data} for a full list of the variables used. 

\begin{figure}
\centering
\begin{scaletikzpicturetowidth}{\textwidth}
\input{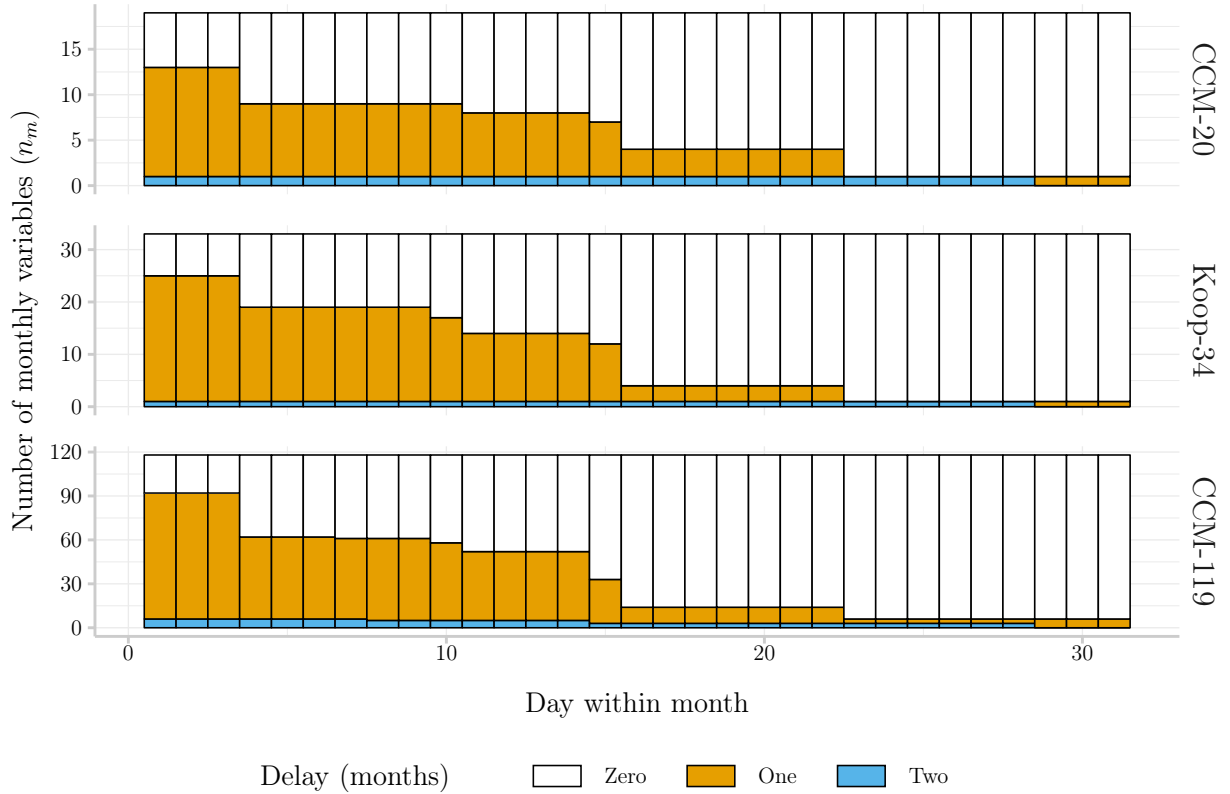}
\end{scaletikzpicturetowidth}
\caption{Structure of the Publication Delay. The height of each bar gives, on a given day, the number of monthly variables for which there is zero, one, or two outcomes missing.}
\label{fig:obs}
\end{figure}

A crucial ingredient in real-time forecasting is the observational pattern in the data. Figure \ref{fig:obs} shows the number of monthly variables with zero, one or two missing observations by day-of-month and by model. At the beginning of the month, a number of variables contain zero missing observations indicating that the outcome of the previous month is already known. These variables include interest rates, spreads and stock market indexes. A few days into the month, new outcomes are obtained. These are predominantly labor market variables. The next major change in observational pattern is around the middle of the month. By this time, most real variables have been observed. However, some of the slow real variables are not observed until the end of the month. As indicated by the bottom blue bar in each panel, some of these are also not observed with a single-month lag, but a two-month lag for the most part.

The quarterly variable GDP growth is assumed to be known in the beginning of the last month in the following quarter; that is, GDP growth for the third quarter is available in the beginning of December.

\subsection{Modeling Stochastic Volatility}
The stochastic volatility model used in this paper is based on the idea of parsimoniously modeling the time-varying volatility. To this end, we use a factor stochastic volatility model and decompose the error term in \eqref{hf} as
\begin{align}
u_t=\Lambda_f f_t+\nu_t,
\end{align}
 where $f_t$ is an $r\times 1$ vector of factors, $\nu_t\sim N(0, \Omega_{t}^\nu)$ and $r\ll n$. Furthermore, $f_t\sim N(0, \Omega_t^f)$ and both $\Omega_t^\nu$ and $\Omega_t^f$ are diagonal with diagonal elements $\omega_{i, t}^\nu$ and $\omega_{j, t}^f$ where the law of motion is
\begin{alignat}{3}
&\log \omega_{i, t}^\nu&& = \mu_i+\phi_i^\nu(\log \omega_{i, t-1}^\nu-\mu_i)+\sigma_i^\nu e^\nu_{i, t}, \quad &&e^\nu_{i, t} \sim N(0, 1)\quad i=1,...n\\
&\log \omega_{j, t}^f &&= \phi_j^f\log \omega_{j, t-1}^f+\sigma_j^fe^f_{j, t}, \quad &&e^f_{j, t} \sim N(0, 1) \quad  j=1,..., r.
\end{alignat} 
The covariance matrix of the VAR innovation is therefore $\Sigma_t=\Lambda_f\Omega_{t}^f\Lambda_f'+\Omega_{t}^\nu$ and thus allows for idiosyncratic stochastic volatilities as well as time-varying covariances by means of the common component. An efficient way of estimating the factor stochastic volatility model was proposed by \cite{Kastner2017}, where the authors made use of the interweaving strategy developed by \cite{Yu2011} and its application to univariate stochastic volatility models by \cite{Kastner2014}. Moreover, \cite{KastnerHuber2018} used the approach in a VAR framework demonstrating the merits of the method.

\subsection{Sampling Regression Parameters}
Conditional on everything but the regression parameters, the original model in \eqref{hf} can be written as
\begin{align}
\tilde{x}_t\equiv x_t - \Lambda_f f_t=\Pi'X_t+\nu_t, \label{eq:xtilde}
\end{align}
where $X_t=(1, x_{t-1}', \dots, x_{t-p}')'$. It is evident from this formulation that the multivariate model consists of $n$ heteroskedastic but independent regressions of $\tilde{x}_{i, t}$ on $X_t$. By letting $\ddot{x}_{i, t}=\frac{1}{\sqrt{\omega^\nu_{i,t}}}\tilde{x}_{i,t}$ and $\ddot{X}_{i, t}=\frac{1}{\sqrt{\omega^\nu_{i,t}}}X_t$ we obtain the factorization
\begin{align}
p(\Pi|\Lambda_f, f, \Omega, x, X)=p(\Pi|\Omega, \tilde{x}, X)=\prod_{i=1}^n p(\Pi^{(i, \cdot)}|\ddot{x}_i, \ddot{X}_i).\label{eq:phidecomp}
\end{align}
In the new parametrization of the model, draws of each row of $\Pi$ can be made independently of each other. This attractive side-effect of the factor stochastic volatility model for sampling the regression parameters was already noted by \cite{KastnerHuber2018}, but we want to stress the benefit thereof. Because of the independence, a parallel implementation of this sampling step is easily employed. \cite{Carriero2019} present a sequential sampling algorithm to sample $\Pi$ row-by-row that dramatically alleviates the issue of sampling from the posterior of $\Pi$, but because it is sequential it is not parallelizable.

Nevertheless, the task of sampling $\Pi$ is---even using parallelization---challenging when dimensions are high. Standard generators of pseudo-random numbers from multivariate normal distributions involve a Cholesky decomposition of the covariance matrix, an operation of order $O(n^3p^3)$ that is to be carried out once per equation. However, because of the special structure of the posterior distribution some improvements can be achieved by using the sampling methods proposed by \cite{Rue2001} or \cite{Bhattacharya2016}, which utilize the specific structure of the posterior. The complexity of the \cite{Bhattacharya2016} sampler is $O(T^2np)$ and tends to be faster than the \cite{Rue2001} algorithm when $np>T$ as the latter has a complexity of $O(n^3p^3)$.  

\section{Simulation Smoothing Using Univariate Adaptive Filtering}
\label{sec:uni}

In order to sample from the posterior of the latent variables, \cite{Schorfheide2015} proposed an alternative representation---called the \emph{compact form}---of the mixed-frequency VAR in which the monthly variables are treated as exogenous. The benefit of leaving the monthly variables out of the state equation is that the computational burden is greatly reduced. However, in real-time forecasting situations when data contain ragged edges, some monthly variables are missing at the end of the sample. This characteristic of the data means that these missing monthly variables will need to be included back in to the state vector. \cite{Schorfheide2015} solved this issue by simply moving from the efficient compact form to the full companion form VAR when the first monthly variable goes missing. As \cite{Ankargren2019} demonstrated, this leads to a large bottleneck when the dimension of the model is high. Consequently, \cite{Ankargren2019} developed an adaptive simulation smoothing algorithm based on the compact form suggested by \cite{Schorfheide2015}. The adaptive part of the algorithm implies that, as some monthly variables go missing at the end of the sample, these are added to the state vector, but not the entire set of monthly variables. By adaptively augmenting the state vector, large efficiency gains can be obtained, particularly when the number of variables and the number of lags are high. For more details, we refer to \cite{Ankargren2019}; the compact and companion forms are also presented in full in Appendix \ref{sec:comp} for self-containment.

Use of the adaptive procedure improves upon filtering of the unbalanced part to the extent that it no longer constitutes the bottleneck of the algorithm when dimensions get higher. It is now instead the filtering step in the compact form that obstructs an efficient procedure.  Fortunately, it is possible to obtain large computational improvements for this chunk of the algorithm by a judicious choice of stochastic volatility model. The method that can be used to further alleviate the procedure of its computational burden is the univariate filtering approach for multivariate time series by \cite{Koopman2000}. Univariate filtering primarily excels when the dimension of the observation vector is large relative to the state vector, which will be the case in the compact form for a large-dimensional mixed-frequency VAR if the number of quarterly variables remains low. The approach is applicable when the state and observation errors are independent and the observation errors feature a diagonal covariance matrix. In what follows, we will discuss how the factor stochastic volatility model allows us to reap the computational benefits of the univariate filtering approach. Since the unbalanced part of the data in the adaptive procedure is responsible for only a negligible part of the computational burden, the discussion of the implementation of the univariate filtering method focuses on the balanced part of the data. There is no obstacle in applying it also on the unbalanced part, but for ease of exposition we do not discuss that any further. It should also be noted that simulation smoothing consists of three main steps: data generation, filtering, and smoothing. We will focus on filtering and smoothing exclusively in the following and refer to \cite{Ankargren2019} for more details on the complete simulation smoothing algorithm.

The \cite{Schorfheide2015} compact formulation of the state-space model for the mixed-frequency VAR is
\begin{equation}
\begin{aligned}
\begin{pmatrix}y_{m,t}\\y_{q,t}\end{pmatrix}&=Z_t\begin{pmatrix}x_{q,t} \\\vdots \\x_{q,t-p}\end{pmatrix}+C_t \begin{pmatrix}y_{m,t-1:t-p}\\ 1\end{pmatrix}+G_t\epsilon_t\\
\begin{pmatrix}x_{q,t} \\\vdots \\ x_{q,t-p}\end{pmatrix}&=T_t\begin{pmatrix}x_{q,t-1} \\\vdots \\x_{q,t-p-1}\end{pmatrix}+D_t\begin{pmatrix}y_{m,t-1:t-p}\\ 1\end{pmatrix}+H_t\epsilon_t,\quad \epsilon_t\sim N(0,I_n)
\end{aligned}\label{eq:compact}
\end{equation}
where the system matrices $\{Z_t, C_t, G_t, T_t, D_t, H_t\}$ are functions of the original model parameters and $y_{m, t-1:t-p}=(y_{m,t-1}', \dots, y_{m, t-p}')'$; see Appendix \ref{sec:comp} for more details.

Since the simulation smoothing step discussed in this section is performed as a Gibbs sampling step, the procedure is carried out conditional on all other parameters in the model. Hence, at this stage $\Lambda_f f_t$ and $\Omega^{\nu}_t$ are known. Partition the factor loadings into monthly and quarterly blocks as $\Lambda_f=(\Lambda_{m, f}', \Lambda_{q, f}')'$ and let
\begin{align}
c_t &= C_t\begin{pmatrix}y_{m,t-1:t-p}\\ 1\end{pmatrix}+\begin{pmatrix}\Lambda_{m,f}f_t \\ 0_{n_q\times1}\end{pmatrix}\\
d_t &= D_t\begin{pmatrix}y_{m,t-1:t-p}\\ 1\end{pmatrix}+\begin{pmatrix}\Lambda_{q,f}f_t \\ 0_{n_qp\times1}\end{pmatrix}.
\end{align}

It is now possible to formulate the compact version of the model as a state-space model with time-varying intercepts and diagonal error covariances:
\begin{align}
y_t&=c_t+Z_t\alpha_t+G_t\epsilon_t\\
\alpha_t&=d_t+T_t\alpha_{t-1}+H_t\epsilon_t,\\
\epsilon_t&\sim N(0, I_n)\\
G_t=\begin{pmatrix}(\Omega_{m, t}^\nu)^{1/2} & 0_{n_m\times n_q} \\ 0_{n_q\times n_m} & 0_{n_q\times n_q}\end{pmatrix},&\quad H_t=\begin{pmatrix}0_{n_q\times n_m} & (\Omega_{q, t}^\nu)^{1/2}  \\ 0_{n_qp\times n_m} & 0_{n_qp\times n_q}\end{pmatrix}.
\end{align}
The univariate filtering method treats the multivariate time series $\{y_t\}$ as a univariate time series. The univariate representation of the observation equation is
\begin{align}
y_{t,i}&=\begin{cases}c_{t,i}+Z_{t,i}\alpha_{t,i}+\sqrt{\omega_{i, t}^\nu}\epsilon_{t,i}, \,& \text{for } i=1, \dots, n_m\\
\begin{pmatrix}0_{1\times n_m} & \Lambda_q\end{pmatrix}\alpha_{t,i}, &\text{for } i=n_m+1, \dots, n_t,\end{cases}\,\,\,\, t=1, \dots, T_b.
\end{align}
 For simplicity, the observation equation does not reflect that $y_{t,i}$ is missing every third period for $i=n_m+1, \dots, n$. Strictly speaking, the observation is empty in these cases. The univariate representation of the state equation is
\begin{align}
\alpha_{t,i+1}&=\alpha_{t,i}, \quad i=1, \dots, n_t-1\\
\alpha_{t+1,1}&=d_t +T_{t+1}\alpha_{t, n_t}+\begin{pmatrix}\epsilon_{t,n_m+1}\\\vdots\\ \epsilon_{t, n} \\ 0_{n_qp\times 1}\end{pmatrix}
\end{align}

The error covariance matrix in the observation equation is diagonal. Univariate filtering exploits the implication of the diagonality in that it brings observations in one at a time instead of bringing the multivariate time $t$ observation in in one go. The computational justification for doing so is that matrix-to-matrix multiplications are avoided in favor of a larger number of scalar or matrix-to-vector multiplications. The latter approach is generally faster if the dimension of the state vector is low in relation to the size of the observation vector.

The standard Kalman filter iterates over the filtering equations to compute $a_t=E(\alpha_t|Y_{t-1})$ and $a_{t|t}=E(\alpha_t|Y_t)$, whereas the univariate approach also computes the intermediate sequence 
\begin{align}
a_{t,i}=E(\alpha_{t,i}|Y_{t-1}, y_{t, 1}, \dots, y_{t,i-1})
\end{align}
for $i=2, \dots, n_t+1$; the connection to the standard filter is $a_{t,1}=a_t$ and $a_{t,n_t+1}=a_{t|t}$. The univariate filtering recursions are, for $t=1, \dots, T_b$ and $i=1, \dots, n_t$:
\begin{alignat}{3}
a_{t,i+1}&=a_{t,i}+K_{t,i}F_{t,i}^{-1}v_{t,i},&\quad P_{t,i+1}&=P_{t,i}-K_{t,i}F_{t,i}^{-1}K_{t,i}'\\
v_{t,i}&=y_{t,i}-Z_{t,i}a_{t,i}-c_{t,i}, &\quad F_{t,i}&=Z_{t,i}P_{t,i}Z_{t,i}'+\omega^\nu_{t, i}, \quad K_{t,i}=P_{t,i}Z_{t,i}',
\end{alignat}
where $Z_{t,i}$ is row $i$ of $Z_t$ and $c_{t,i}$ is the $i$th element of $c_t$.
To transition from time $t$ to time $t+1$, the following relations are used:
\begin{align}
a_{t+1, 1}&=T_{t+1} a_{t, n_t+1}+d_{t+1}, \quad P_{t+1,1}=T_{t+1}P_{t, n_t+1}T_{t+1}'+H_{t+1}H_{t+1}'.
\end{align}
Several of the Kalman filter's familiar terms---such as $F_t$---now appear as scalars instead of matrices. The consequence is that there is no need to invert the full $F_t$ matrix; instead, the only inversion required is that of the scalar $F_{t,i}$. 

Finally, to conclude the description of the univariate filtering approach, the recursions used for conducting the smoothing step of the algorithm are:
\begin{equation}
\begin{aligned}
L_{t,i}&=I_{n_q(p+1)}-K_{t,i}Z_{t,i}F_{t,i}^{-1}\\
r_{t,i-1}&=Z_{t,i}'F_{t,i}^{-1}v_{t,i}+L_{t,i}'r_{t,i}\\
r_{t-1, p}&=T_t'r_{t,0}
\end{aligned}
\label{eq:unismooth}
\end{equation}
for $i=n_t, \dots, 1$ and $t=T_b, \dots, 1$. The smoothed estimate of primary interest is computed as $\hat{\alpha}_t=a_t+P_tr_{t,0}$.

\subsection{Quantifying the Computational Improvement}
\label{sec:quant}
While the number of variables in the model determines the computational burden to a large extent, the choice of the lag length $p$ naturally plays a crucial role. In the large VAR literature, the choice of lag length is subject to a notable degree of variation. For example, \cite{KastnerHuber2018} used $p=1$, whereas \cite{Schorfheide2015} and \cite{Gotz2018} for mixed-frequency VARs set $p=6$, and \cite{Banbura2010,Carriero2019} let $p=13$. Studying VAR specifications more systematically, \cite{Carriero2015b} found that optimizing the lag length improved the forecasting ability as compared to a baseline specification using $p=13$, although only by a few percentage points. Therefore, the conclusion one can draw is that lag lengths used in practice are relatively idiosyncratic, and that choosing a large lag length is a way to play it safe.

\begin{figure}
\centering
\begin{scaletikzpicturetowidth}{\textwidth}
\input{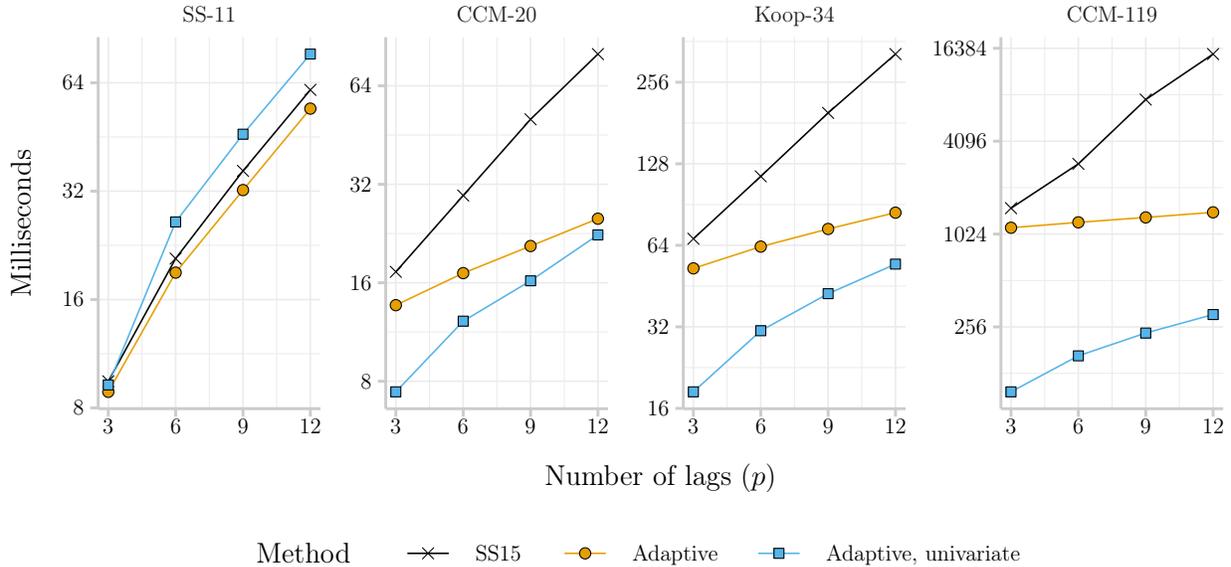}
\end{scaletikzpicturetowidth}
\caption{Computational Complexities of Simulation Smoothers. The adaptive algorithms scale better with $p$, and the benefit of the univariate algorithm grows with model size.}
\label{fig:timing_data}
\end{figure}

Choosing a large lag length in the mixed-frequency VAR context is more expensive than if data are all on the same frequency, since the dimension of the state vector is $n_q(p+1)$ and thereby enlarged when $p$ increases. Moreover, the more quarterly variables that are included (for a fixed $n$), the heavier the computational burden. Figure \ref{fig:timing_data} shows the cost of increasing the lag length for the mixed-frequency block of the algorithm for the three models discussed in Section \ref{sec:application} with $n=20,34$ and 119, respectively, and a single quarterly variable. In addition, the figure displays the cost of increasing $p$ for the \citet[SS-11 in the figure]{Schorfheide2015} model, which uses eight monthly and three quarterly variables. Of the eight monthly variables, three are available with no publication delay and five with a one-month delay. For the CCM-20, Koop-34 and CCM-119 models we use the observational pattern from day 15, see Figure \ref{fig:obs}.

Figure \ref{fig:timing_data} shows that our adaptive algorithm with univariate filtering works well for the empirical models when there is a single quarterly variable. The increased computational effort that is needed for a higher lag length is modest, whereas the \cite{Schorfheide2015} algorithm does not scale well. The first panel of the figure shows that when there are three quarterly variables, the adaptive algorithm offers an improvement compared with the \cite{Schorfheide2015} procedure, while combining it with univariate filtering performs worse. The inferior performance of univariate filtering is not very surprising---the method primarily excels when the dimension of the state vector is small relative to the dimension of the observation vector. For $p>3$, the state vector is larger than the observation vector. In contrast, the state vector is always smaller than the observation vector for the other three models. Comparing the relative performances across model sizes also reveals that when $n$ increases, the relative gains of univariate filtering increase. The reason is that $n_q(p+1)$ as a share of $n$ decreases, which further pushes univariate filtering into its domain of superiority. In particular, when the model includes 119 variables (rightmost panel), the difference in running time between the adaptive algorithm with univaraite filtering and the \cite{Schorfheide2015} method is substantial---almost 16 seconds per iteration is required by the latter, as opposed to 0.4 seconds.

To summarize, there are two key points made by Figure \ref{fig:timing_data}. First, the gains from using univariate filtering from a computational perspective become larger as the dimension of the model increases. Second, varying the lag length is relatively cheap using either the adaptive or the adaptive with univariate filtering algorithms, which thereby allow for larger freedom in selecting the number of lags.

\subsection{Other Stochastic Volatility Models}

The preceding discussion of how the univariate filtering approach can be used for the balanced part of the sample focused on the case when a factor stochastic volatility model is used to model time-varying variances in the model. This is by no means the only way to model the stochastic volatility component. A popular alternative is to decompose $\Sigma_t=A_t^{-1}\Lambda_t\left(A_t^{-1}\right)'$ where $A_t$ is a lower triangular matrix and $\Lambda_t$ is a diagonal matrix. In a seminal paper, \cite{Primiceri2005} let the non-zero elements of $A_t$ as well as the log of the diagonal of $\Lambda_t$ evolve as driftless random walks, whereas \cite{Cogley2005} let $A_t$ be constant over time. Among the literature close to ours, \cite{Clark2011} followed \cite{Cogley2005} using a constant $A_t$ while \cite{Cimadomo2016} let it too vary over time. In order to model larger systems, \cite{Carriero2016} proposed the use of a common drifting volatility where $\Sigma_t=f_t\Sigma$ and $f_t$ is a scalar geometric random walk. Thus, a change in volatility for the entire $n$-dimensional system is achieved by scaling the constant covariance matrix $\Sigma$ by $f_t$. This line of modeling was later also pursued in the context of mixed-frequency VARs by \cite{Gotz2018,Ankargren2018}. In related work, \cite{Carriero2019} propose a sampling procedure which greatly mitigates the computational burden in large VARs with $\Sigma_t=A^{-1}\Lambda_t\left(A^{-1}\right)'$ by exploiting the triangular structure of $A$.

We will focus next on two of the aforementioned stochastic volatility models: the common drifting volatility model used by \cite{Carriero2016,Gotz2018} and the model used by \cite{Cogley2005}. The latter model uses $\Sigma_t=A^{-1}\Lambda_t\left(A^{-1}\right)'$ where the Cholesky factor $L_tL_t'=\Sigma_t$ is $L_t=A^{-1}\Lambda_t^{0.5}$. For the common drifting volatility model, we can let $\Sigma=P^{-1}D(P^{-1})'$ where $P$ is lower diagonal with ones on the main diagonal and $D$ is a diagonal matrix. If we let $D_t=f_tD$, the common drifting volatility model can be written as $\Sigma_t=P^{-1}D_tP^{-'}$, i.e. the same form as the \cite{Cogley2005} model. In the following, it is therefore, for our purposes, enough to deal with the $\Sigma_t=A^{-1}\Lambda_t \left(A^{-1}\right)'$ case as this encompasses the common drifting stochastic volatility model.

To transform the model into a form that enables the use of the univariate filtering approach, it is necessary to transform the monthly VAR model such that the errors are independent:
\begin{align}
x_t^*&=\sum_{j=1}^p\Pi_j^* x_{t-j}^*+\Lambda_t^{0.5}e_t,\quad e_t\sim N(0, I_n)
\end{align}
where $x_t^*=Ax_t$ and $\Pi_j^*=A\Pi_jA^{-1}$. The observation equation is similarly adjusted to account for the transformation:
\begin{align}
y_t^*=S_t\Lambda^*\begin{pmatrix}x_t^*\\\vdots\\x_{t-p+1}^*\end{pmatrix},
\end{align}
where $y^*_t=Ay_t$ and $\Lambda^*=A\Lambda (I_p\otimes A^{-1})$. However, this seemingly innocuous transformation comes with one drawback: for most choices of aggregation matrices $\Lambda$, the rows of $\Lambda^*$ corresponding to quarterly variables will consist of only non-zero elements. The consequence of this is that $x_{q,t}^*$ depends concurrently on $x_{m, t}^* $---a relation that is not predetermined at time $t$ and hence cannot be moved to the exogenous component of the model. It would therefore be necessary to keep $x_{m,t}^*$---although not its lags---in the state vector of the compact form, thereby creating a state vector of dimension $(n+n_qp)\times1$. Differently put, for these stochastic volatility models, enabling the use of the univariate filtering procedure comes at the expense of requiring an additional $n_m$ terms to be put into the state vector. However, the adaptive algorithm proposed by \cite{Ankargren2019} is still applicable and may provide considerable computational improvements relative to the original \cite{Schorfheide2015} procedure as demonstrated by Figure \ref{fig:timing_data}.
\section{Empirical Application: Estimating Large Mixed-Frequency VARs for the US}
\label{sec:crisis}

In this section, we estimate the mixed-frequency VARs with factor stochastic volatility using the three sets of data discussed in Section \ref{sec:application}. We estimate the three models on the full data and discuss some of the key features.

\subsection{Implementation Details}
The sample we use for estimation starts in January 1980. We follow previous work that has used mixed-frequency VARs for US data, see e.g. \cite{Schorfheide2015,Gotz2018}, and use $p=6$ lags. For the number of factors, \cite{KastnerHuber2018} found that a factor structure with two factors was preferable in terms of joint log predictive score for a selected subset of variables using their quarterly model with 215 variables, while one or no factor was preferable with respect to univariate density forecasting of GDP. Thus, since their results suggest zero to two factors for a substantially larger model than ours, we settle on a single $r=1$ factor in all three of our models. Each model is estimated using 30,000 MCMC draws, where the initial 10,000 are discarded for burn-in. Of the remaining 20,000, we save every 20$^{\text{th}}$ draw to improve mixing and reduce storage costs. It can be noted that the CCM-119 model draws $n(np+1)=1$85,085 regression parameters and $T(n+r)=55,320$ volatilities at every iteration and therefore requires a substantial amount of storage, which makes thinning a necessity. We leverage multiple cores to speed up our computations and draw regression parameters in parallel. Using an Intel Xeon E5-2630 v4 processor with 10 cores, roughly half a second per draw is needed for CCM-119 when estimated on the full sample. While computational resources, models and implementations are inevitably different, half a second per draw can be contrasted to the work by \cite{Carriero2019} who document roughly five seconds per draw. Therefore, by current standards the computational burden of our large mixed-frequency VAR is undoubtedly within the realms of acceptability.

\subsection{Mixing of MCMC Samplers}

Thinning the draws has, in addition to mitigating storage costs, the advantage of reducing the correlation between the draws that are stored, thereby improving the share of information contained in the draws saved. To evaluate the mixing of an MCMC sampler, a common metric is the inefficiency factor (see e.g. \citealp{Chib2001}). The inefficiency factor for the $i$th parameter is
\begin{align}
IF(i)=1+2\sum_{j=1}^\infty\rho_{i, j},
\end{align}
where $\rho_{i,j}$ is the $j$th autocorrelation of the $i$th parameter. A value of one is obtained when there is no autocorrelation. Consequently, values close to one indicate low autocorrelation in the MCMC sampler and a large degree of information in the samples. Large values, on the other hand, imply that the chain contains little information due to high autocorrelation. A value below 20 is often used as a rule of thumb for good mixing \citep{Primiceri2005}. We estimate the inefficiency factor using the \texttt{coda} package for \texttt{R} \citep{Plummer2006}. Table \ref{tab:ifs} provides a summary of the distributions of inefficiency factors by group of parameter.

\begin{table}[t]
\small 
\caption{\label{tab:ifs}Inefficiency Factors}
\centering
\begin{tabular}{ccccccccc}
\toprule
&&&\multicolumn{4}{c}{Percentiles}\\
\cmidrule{4-7}
 Model & \# of pars & Min & 50$^{\text{th}}$ & 75$^{\text{th}}$ & 95$^{\text{th}}$ & 99$^{\text{th}}$ & Max & \%$ > 20$\\
\midrule
\addlinespace[0.3em]
\multicolumn{9}{l}{\textbf{Latent monthly GDP}}\\
\hspace{0.25em} CCM-20 & 460 & 0.5 & 1.2 & 1.5 & 2.2 & 3.0 & 4.3 & 0.0\\

\hspace{0.25em} Koop-34 & 414 & 0.9 & 1.4 & 1.9 & 3.5 & 4.8 & 7.0 & 0.0\\

\hspace{0.25em} CCM-119 & 460 & 1.0 & 9.0 & 13.4 & 23.8 & 35.2 & 66.3 & 7.8\\

\addlinespace[0.3em]
\multicolumn{9}{l}{\textbf{Regression parameters}}\\
\hspace{0.25em} CCM-20 & 2420 & 0.5 & 1.0 & 1.0 & 2.4 & 10.0 & 50.1 & 0.3\\

\hspace{0.25em} Koop-34 & 6970 & 0.3 & 1.0 & 1.0 & 1.3 & 1.9 & 11.2 & 0.0\\

\hspace{0.25em} CCM-119 & 85085 & 0.3 & 1.0 & 1.0 & 1.2 & 2.0 & 82.5 & 0.0\\

\addlinespace[0.3em]
\multicolumn{9}{l}{\textbf{Latent factor}}\\
\hspace{0.25em} CCM-20 & 460 & 1.0 & 8.6 & 11.0 & 14.4 & 17.6 & 45.1 & 0.7\\

\hspace{0.25em} Koop-34 & 414 & 0.8 & 2.4 & 3.0 & 4.2 & 6.0 & 10.3 & 0.0\\

\hspace{0.25em} CCM-119 & 460 & 3.2 & 18.2 & 31.7 & 72.0 & 104.6 & 139.7 & 45.0\\

\addlinespace[0.3em]
\multicolumn{9}{l}{\textbf{Factor loadings}}\\
\hspace{0.25em} CCM-20 & 20 & 1.0 & 3.0 & 4.5 & 9.9 & 11.8 & 12.3 & 0.0\\

\hspace{0.25em} Koop-34 & 34 & 1.0 & 1.8 & 3.1 & 5.7 & 6.7 & 7.0 & 0.0\\

\hspace{0.25em} CCM-119 & 119 & 0.7 & 1.4 & 2.8 & 18.6 & 38.9 & 321.4 & 4.2\\

\addlinespace[0.3em]
\multicolumn{9}{l}{\textbf{Latent log-volatilities}}\\
\hspace{0.25em} CCM-20 & 9660 & 0.4 & 1.0 & 1.2 & 4.7 & 13.9 & 21.6 & 0.0\\

\hspace{0.25em} Koop-34 & 14490 & 0.4 & 1.0 & 1.2 & 2.0 & 5.7 & 22.3 & 0.0\\

\hspace{0.25em} CCM-119 & 55200 & 0.3 & 1.0 & 1.2 & 1.9 & 12.4 & 65.5 & 0.6\\

\addlinespace[0.3em]
\multicolumn{9}{l}{\textbf{Means of log-volatilities}}\\
\hspace{0.25em} CCM-20 & 20 & 1.0 & 1.0 & 1.2 & 3.9 & 15.4 & 18.3 & 0.0\\

\hspace{0.25em} Koop-34 & 34 & 0.9 & 1.3 & 1.7 & 4.2 & 7.8 & 9.1 & 0.0\\

\hspace{0.25em} CCM-119 & 119 & 1.0 & 1.2 & 1.5 & 3.1 & 5.9 & 15.7 & 0.0\\

\addlinespace[0.3em]
\multicolumn{9}{l}{\textbf{AR parameters for log-volatilities}}\\
\hspace{0.25em} CCM-20 & 21 & 1.4 & 3.2 & 4.2 & 6.6 & 6.6 & 6.6 & 0.0\\

\hspace{0.25em} Koop-34 & 35 & 1.3 & 3.3 & 4.1 & 5.4 & 12.9 & 16.7 & 0.0\\

\hspace{0.25em} CCM-119 & 120 & 1.2 & 2.8 & 3.9 & 5.8 & 9.3 & 10.8 & 0.0\\

\addlinespace[0.3em]
\multicolumn{9}{l}{\textbf{Innovation variances for log-volatilities}}\\
\hspace{0.25em} CCM-20 & 21 & 1.5 & 3.1 & 5.2 & 7.3 & 21.2 & 24.6 & 4.8\\

\hspace{0.25em} Koop-34 & 35 & 1.8 & 3.2 & 4.1 & 6.0 & 12.9 & 16.2 & 0.0\\

\hspace{0.25em} CCM-119 & 120 & 2.0 & 3.3 & 4.6 & 7.2 & 12.1 & 15.2 & 0.0\\
\bottomrule
\end{tabular}
\end{table}

Table \ref{tab:ifs} shows that the inefficiency factors for CCM-20 and Koop-34 are generally well below or around 20, with the exception of a handful of outliers. The vast majority of parameters in the CCM-119 model also display satisfactory mixing, but the size of the outliers is somewhat larger. The most problematic aspect of the model is the mixing of the latent factor. It is encouraging, however, that mixing is excellent for the other parameters in the model. The approach we have discussed in this paper enables us to speed up computations and, for a fixed amount of time, produce more draws from the posterior distribution. We have noticed that mixing of the the latent high-frequency series can be troublesome if the number of draws is limited and so while other methods for large VARs can be extended to handle mixed-frequency data, the benefit of the route taken in this paper is that we can obtain draws of higher quality in the same amount of time.

\subsection{Results}

\begin{figure}
\centering
\begin{scaletikzpicturetowidth}{\textwidth}
\input{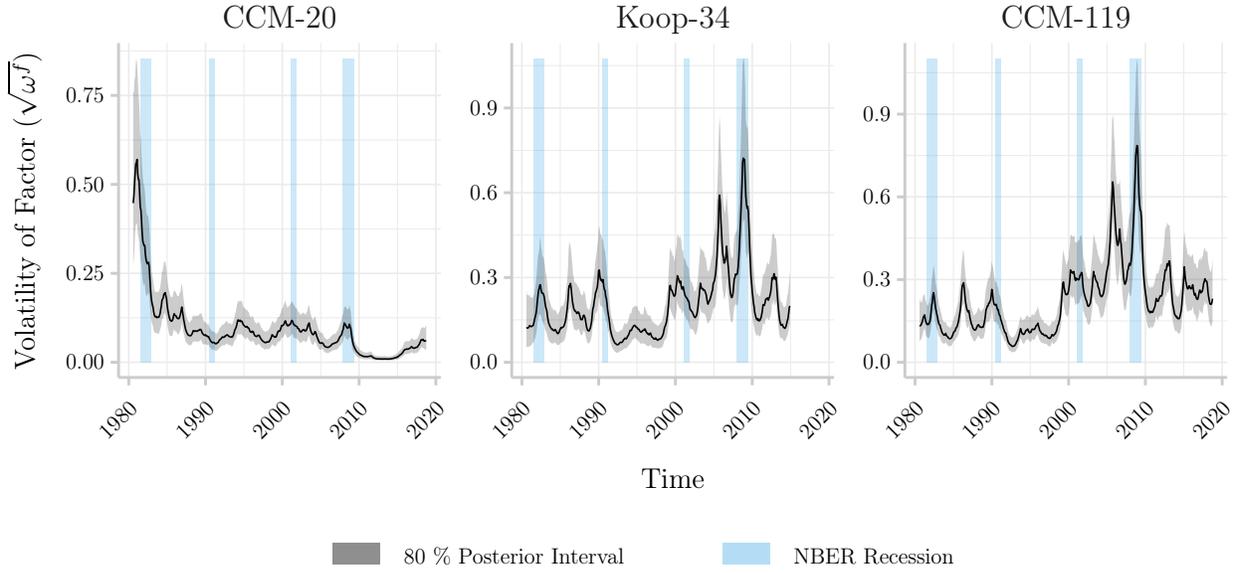}
\end{scaletikzpicturetowidth}
\caption{Posterior Volatility of Latent Factor (Standard Deviation). The figure shows the median, and the 10$^\text{th}$ and 90$^\text{th}$ percentiles of the simulated posterior distribution of the square root of the volatility of the factor ($\sqrt{\omega^f}$). The vertical shading indicates periods of recession as defined by the National Bureau of Economic Research. The sample used for estimating the Koop-34 model ends in late 2014 due to the discontinuation of some of the included series.}
\label{fig:facvol}
\end{figure}

We now turn to the estimated models and illustrate some of their core features. To begin, we study the volatility of the latent factor. Figure \ref{fig:facvol} plots the volatilities estimated by each model. To simplify the interpretation, the figure includes periods of contraction as defined by the National Bureau of Economic Research.

Figure \ref{fig:facvol} shows a remarkable similarity between the factor volatilities estimated by Koop-34 and CCM-119, with major peaks around all recessions in the sample. The volatility estimated by CCM-20, however, is quite different as it estimates the volatility of the factor to be substantially higher in the beginning of the sample during the 1981--1982 recession. Nevertheless, the volatility in the CCM-20 model did show signals of increased volatility during the most recent two contractions. That volatility is heightened is more evident in the volatilities estimated by the Koop-34 and CCM-119 models, where all recessions coincide with peaks in volatility. 

\begin{figure}
\centering
\begin{scaletikzpicturetowidth}{0.9\textwidth}
\input{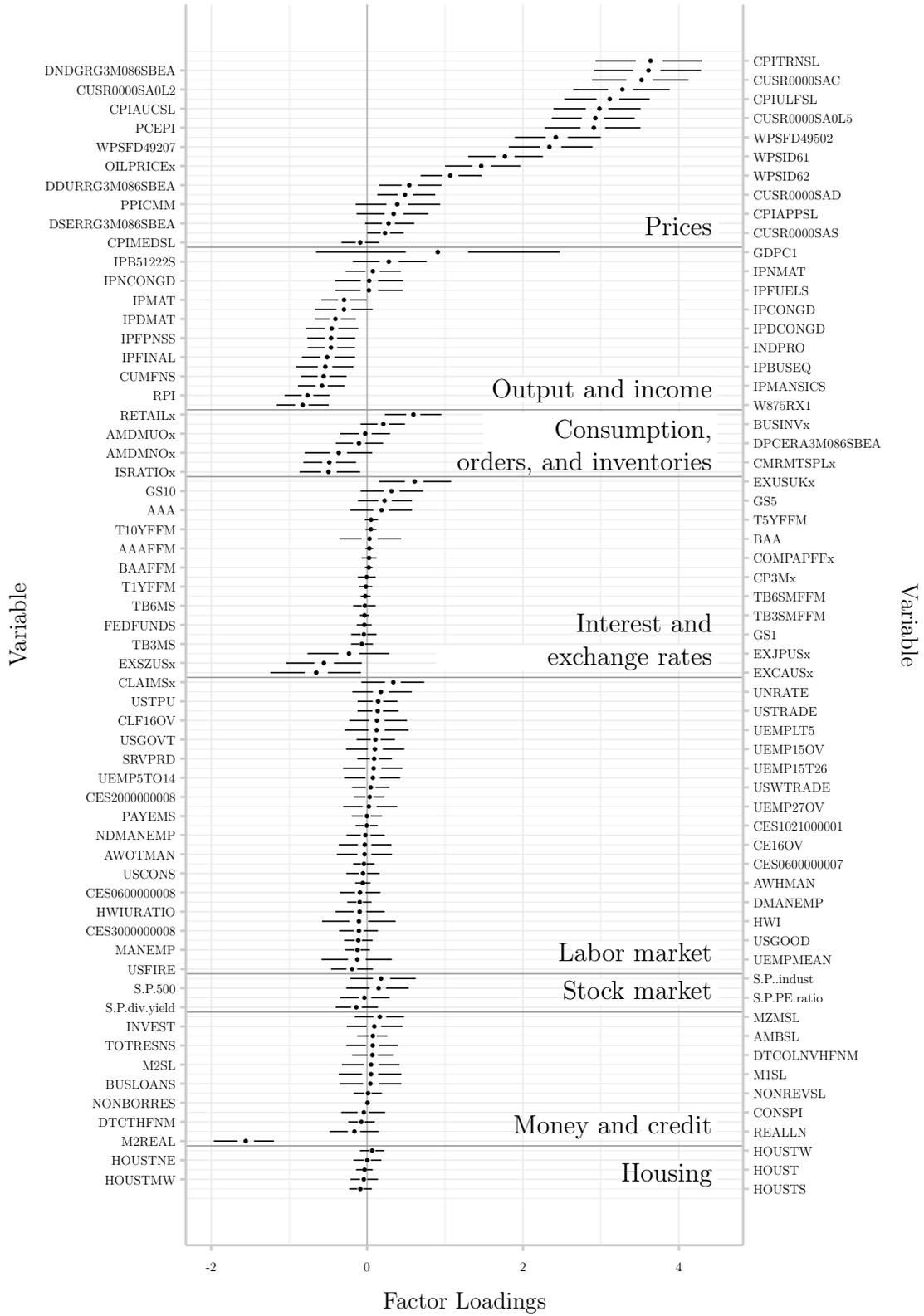}
\end{scaletikzpicturetowidth}
\caption{Boxplots of Posterior Factor Loadings in the CCM-119 Model. Points represent medians, and lines extend from the upper (lower) quartile to the maximum (minimum) value within 1.5 times the interquartile range.}
\label{fig:facload}
\end{figure}

In order to better understand the meaning of the factor and its volatility, Figure \ref{fig:facload} displays boxplots of the posterior distribution of the factor loadings in the CCM-119 model.\footnote{Without further restrictions, the sign of the vector of loadings is not identified. We identify the sign of the vector of loadings a posteriori using the ``maximin'' method suggested by \cite{Kastner2017}. The sign at every iteration is selected such that the loading that has the largest minimum of absolute values of draws is positive at every iteration.}  From the figure, it is evident that the factor largely captures comovements in prices as most of the variables in this group are associated with large posterior loadings. Moreover, most of the variables in the output and income category load on the factor while the volatility of, e.g., labor market and stock market variables are mainly driven by their respective idiosyncratic components. We can also see that the posterior distribution of the GDP loading is wider than for other loadings. It is hardly surprising that we find a wider posterior distribution for the GDP loading as the factor, and hence the associated loadings, relate to the monthly frequency, at which the GDP series changes from iteration to iteration.

\begin{figure}
\centering
\begin{scaletikzpicturetowidth}{\textwidth}
\input{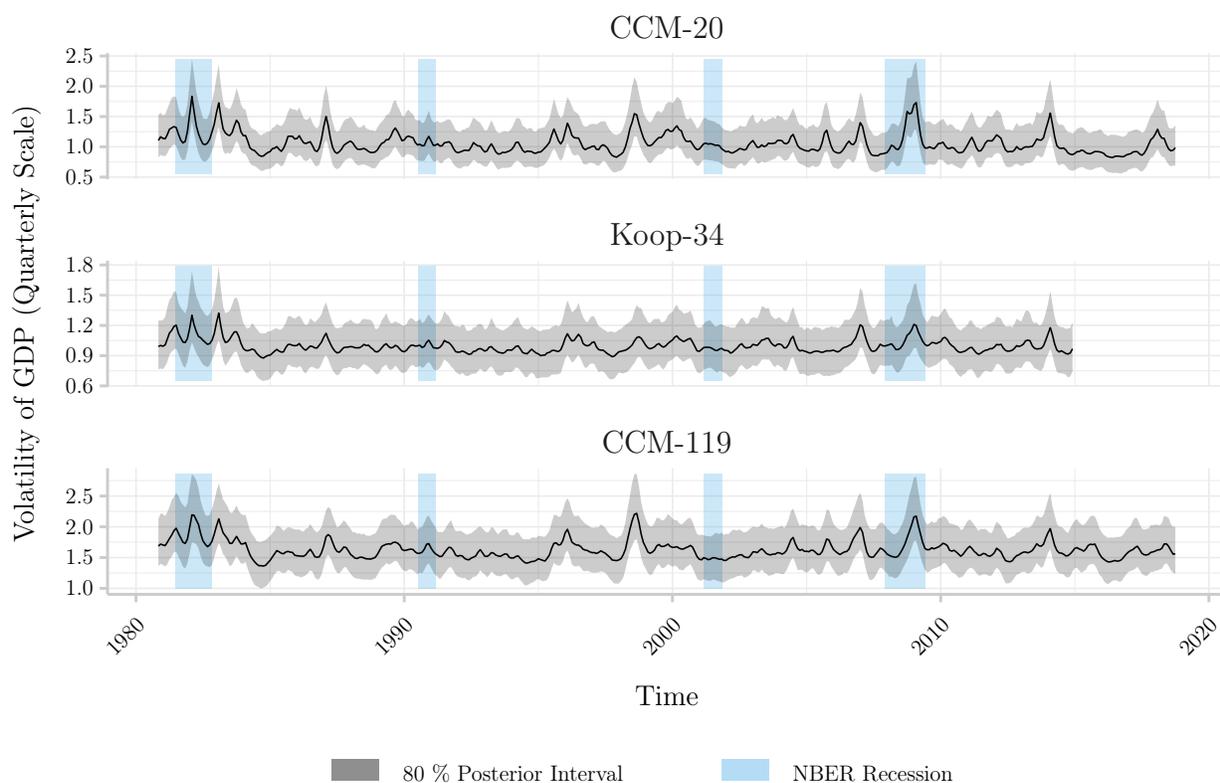}
\end{scaletikzpicturetowidth}
\caption{Posterior Volatility of Quarterly GDP (Standard Deviation). The figure shows the median, and the 10$^\text{th}$ and 90$^\text{th}$ percentiles of the simulated posterior distribution of the square root of the implied volatility for GDP. The monthly volatility is transformed to the quarterly scale by using the triangular aggregation scheme, and the plot shows quarterly volatility for time periods when GDP was observed. The vertical shading indicates periods of recession as defined by the National Bureau of Economic Research. The sample used for estimating the Koop-34 model ends in late 2014 due to the discontinuation of some of the included series.}
\label{fig:gdpvol}
\end{figure}

As a final illustration of the model's features, we next study the implied volatility of GDP. It follows from the construction of the model that the conditional variance of the error term in the equation for monthly GDP is
\begin{align}
\sigma^2_{t, GDP}=V(u_{GDP, t}|\Lambda_f, \Omega)=\lambda_{GDP, f}^2\omega_t^f+\omega_{GDP,t}^\nu,
\end{align}
where $\lambda_{GDP, f}$ is the GDP loading. In order to improve interpretability and make the volatility more comparable to the observed outcomes of GDP, we compute and plot ``aggregated'' volatility where $\sigma^2_{t, GDP}$ is aggregated using the triangular aggregation scheme discussed in Section \ref{sec:background}. Figure \ref{fig:gdpvol} shows the volatilities over time.

While the volatility of the factor in Figure \ref{fig:facvol} differed substantially between the small CCM-20 model and the other two models, Figure \ref{fig:gdpvol} shows that the volatility of GDP itself is strikingly similar across models. The figure reveals that the recessions in the 1990s and early 2000s largely had no implications for the volatility of GDP, whereas the crises in the 1980s and late 2000s had major effects on GDP volatility. 

\section{Concluding Remarks}
\label{sec:conclusion}
We have in this paper developed a Bayesian vector autoregressive model with stochastic volatility that is suitable for large-dimensional mixed-frequency datasets. Our particular interest lies in leveraging a large set of monthly variables for modeling quarterly GDP growth. The model is made feasible for large dimensions by assuming a factor stochastic volatility structure. The factor stochastic volatility model enables us to refine the adaptive simulation smoothing algorithm developed by \cite{Ankargren2019} by coupling their adaptive algorithm with the univariate filtering technique proposed by \cite{Koopman2000}. The key feature making this possible is that the factor stochastic volatility model renders the equations conditionally independent. The conditional independence among the equations of the VAR also allows us to draw the regression parameters on an equation-by-equation basis in parallel. 

The features of the model are illustrated by estimation on three datasets, consisting of $n=20$, $n=34$ and $n=119$ variables, respectively. The three datasets all contain monthly data as well as quarterly GDP growth. We extract the volatility of the common component and find that the volatility is largely in line with most expectations, with heightened volatility around periods of recession. In terms of volatility of GDP growth itself, the three models deliver similar estimates with major peaks in the early 1980s and late 2000s in particular, with other contractions having played smaller roles for the volatility of GDP. Analyzing the factor loadings of the large CCM-119 in some more detail, we find that the factor predominantly captures co-movements of variables in the prices and output and income sectors.

We also investigate the mixing of the MCMC samplers that we use for estimating the models. Generally, our three models mix reasonably well after thinning given their size. However, mixing becomes slower as the dimensions increase, and a larger number of draws is needed in order to produce the same amount of information due to the correlation among draws. An advantage of the procedure that we propose is that draws are generally cheaper due to our computationally more efficient simulation smoother and our ability to draw regression parameters in parallel. These benefits cannot be enjoyed if other stochastic volatility models are used. The most problematic aspect of the MCMC sampler is the latent monthly GDP series and the latent factor. An interesting topic for further exploration of the model would be to sample these two latent variables jointly. The simulation smoother that is used for sampling latent monthly GDP can easily be augmented with the latent factor, which would enable joint sampling that may be helpful for the quality of the MCMC procedure. The caveat is that the observation equation error covariance matrix will no longer be diagonal, and univariate filtering is not applicable any more and the computational burden is increased. Moreover, the sample size $T$ plays an important role for the computational complexity of all of the major Gibbs sampling blocks: larger $T$ means that simulation smoothing needs to be carried out for a longer sample, which affects both the sampling of latent monthly variables, volatilities and factors. Additionally, a larger $T$ increases the cost of sampling the regression parameters when the model is truly high-dimensional so that $T<np$. When $T<np$, the regression parameters can more efficiently be sampled using the \cite{Bhattacharya2016} sampler, whose complexity is $O(T^2np)$ as opposed to the \cite{Rue2001} sampler that we use otherwise, which has complexity $O(n^3p^3)$. The issue of computational time when $T$ is large is not unique to the mixed-frequency model, although it is arguably more pronounced, but is also explicitly discussed by e.g. \cite{KastnerHuber2018}. 

A promising avenue for future research is developing computational strategies similar to ours that allow speed enhancements also for other stochastic volatility models. When the number of variables is in the hundreds and the lag length goes beyond single-digit numbers, the speed-ups we obtain from univariate filtering and parallelization make a considerable difference. Finding approaches that yield these types of improvements for other types of stochastic volatility models would enrichen the class of feasible mixed-frequency models for large dimensions and would be highly valuable contributions. Furthermore, a thorough forecasting evaluation including our suggested model and other contenders for forecasting in the presence of large mixed-frequency datasets should be of interest for many macroeconomic forecasters.

\bibliographystyle{apalikedoi}
\bibliography{refs_fsv}

\clearpage
\appendix

\section{The Compact and Companion Forms}
\label{sec:comp}

The companion form is obtained by stacking \eqref{hf} and casting in the form a VAR(1), i.e.
\begin{align}
\begin{pmatrix}
x_t \\ x_{t-1}\\ \vdots \\ x_{t-p+1}
\end{pmatrix}
=\begin{pmatrix}
\Pi_c \\ 0 \\ \vdots \\ 0
\end{pmatrix}
+
\begin{pmatrix}
\Pi_1 & \Pi_2 & \cdots &\Pi_{p-1} & \Pi_p\\
I_n & 0 & \cdots & 0 & 0 \\
\vdots & \vdots & \ddots & \vdots \\
0 & 0 & \cdots & I_n & 0
\end{pmatrix}
\begin{pmatrix}
x_{t-1} \\ x_{t-2} \\ \vdots \\ x_{t-p}
\end{pmatrix}
+
\begin{pmatrix}
u_t  \\ 0 \\ \vdots \\ 0
\end{pmatrix}.\label{eq:statecompanion}
\end{align}
Equation \eqref{eq:statecompanion} is the state equation in the companion form VAR, and collecting \eqref{eq:obseq} in
\begin{align}
y_t=\begin{pmatrix}S_{m,t} & 0 \\ 0 & S_{q, t}\end{pmatrix}\Lambda \begin{pmatrix}
x_t \\ x_{t-1}\\ \vdots \\ x_{t-p+1}
\end{pmatrix}
\end{align}
yields the observation equation, where $\Lambda$ is the aggregation matrix. 

The compact form of the model is displayed in \eqref{eq:compact}. Let $\Pi_{mq}$ be the $n_m\times pn_q$ matrix obtained by selecting the first $n_m$ rows of $(\Pi_1, \dots, \Pi_p)$ and the columns that correspond to lagged values of the quarterly variables. The submatrix $\Pi_{mq}$ then contains the regression parameters from the equations for the monthly variables that account for lags of the quarterly variables. Similarly, let $\Pi_{qm}$ be the $n_q\times pn_m$ matrix with parameters that account for the $p$ lags of the $n_m$ variables in the $n_q$ equations for the quarterly variables. Finally, let $\Pi_{mm}$ and $\Pi_{qq}$ be the $n_m\times pn_m$ and $n_q\times pn_q$ matrices containing parameters for the lagged effects of monthly (quarterly) variables in the equations for the monthly (quarterly) variables. Then the system matrices in the compact form state-space model are
\begin{alignat}{3}
Z_t =& \begin{pmatrix}
0_{n_m\times n_q} & \Pi_{mq}\\
\Lambda_{qq} & 0_{n_q\times n_q}
\end{pmatrix}, & \quad T_t =& \begin{pmatrix}
 \Pi_{qq} & 0_{n_q\times n_q}\\
 I_{pn_q}& 0_{pn_q\times n_q}
\end{pmatrix},\\
C_t =& \begin{pmatrix}
\Pi_{mm} & \Pi_{mc}\\
0_{n_q\times  pn_m} & 0_{n_q\times 1}
\end{pmatrix}, &\quad D_t =& \begin{pmatrix}
\Pi_{qm} & \Pi_{qc}\\
0_{pn_q\times  pn_m} & 0_{pn_q\times 1}
\end{pmatrix}\\
G_t  =& \begin{pmatrix}
(\Sigma_t^{1/2})_{m, \bullet}\\
0_{n_q\times n}
\end{pmatrix}, &\quad
H_t  =& \begin{pmatrix}
(\Sigma_t^{1/2})_{q, \bullet}\\
0_{pn_q\times n},
\end{pmatrix}
\end{alignat}
where $\Sigma_t^{1/2}$ is the lower triangular Cholesky factor of $\Sigma_t$ and 
\begin{align}
\Sigma_t^{1/2}=\begin{pmatrix}(\Sigma_t^{1/2})_{m,\bullet} \\ (\Sigma_t^{1/2})_{q,\bullet}\end{pmatrix}.
\end{align}

The univariate procedure that is discussed in Section \ref{sec:uni} is used within the adaptive algorithm suggested by \cite{Ankargren2019}. It replaces the filtering and smoothing steps in the compact form, with the steps for the unbalanced part of the data kept intact. See \cite{Ankargren2019} for more details and the algorithm presented in full.
\clearpage
\section{Data}
\label{sec:data}

{\scriptsize
\begin{longtable}{lrrccc}
\caption{Publication Information and Series ID of Data. The ID column gives the FRED-MD \citep{McCracken2016} series ID for each series. The Month and Day columns display the number of months and days after a month has ended until the new datum is published (i.e., 1 and 23 means that the datum was published on the 23rd of the subsequent month). In the final three columns, X indicates inclusion in the model.}\\

\toprule
ID & Month & Day & CCM-20 & Koop-34 & CCM-119\\
\midrule
\endfirsthead

\multicolumn{6}{c}%
{{\bfseries \tablename\ \thetable{} -- continued from previous page}} \\
\toprule
ID & Month & Day & CCM-20 & Koop-34 & CCM-119\\
\midrule
\endhead

\hline \multicolumn{6}{r}{{Continued on next page}} \\ \hline
\endfoot

\bottomrule
\endlastfoot
RPI & 1 & 23 & X & X & X\\
W875RX1 & 1 & 23 &  &  & X\\
INDPRO & 1 & 16 & X & X & X\\
IPFPNSS & 1 & 16 &  &  & X\\
IPFINAL & 1 & 16 &  &  & X\\
\addlinespace
IPCONGD & 1 & 16 &  &  & X\\
IPDCONGD & 1 & 16 &  &  & X\\
IPNCONGD & 1 & 16 &  &  & X\\
IPBUSEQ & 1 & 16 &  &  & X\\
IPMAT & 1 & 16 &  &  & X\\
\addlinespace
IPDMAT & 1 & 16 &  &  & X\\
IPNMAT & 1 & 16 &  &  & X\\
IPMANSICS & 1 & 16 &  &  & X\\
IPFUELS & 1 & 16 &  &  & X\\
CUMFNS & 1 & 16 & X & X & X\\
\addlinespace
CLF16OV & 1 & 4 &  &  & X\\
CE16OV & 1 & 4 &  &  & X\\
UNRATE & 1 & 4 & X & X & X\\
UEMPMEAN & 1 & 4 &  &  & X\\
UEMPLT5 & 1 & 4 &  &  & X\\
\addlinespace
UEMP5TO14 & 1 & 4 &  &  & X\\
UEMP15OV & 1 & 4 &  &  & X\\
UEMP27OV & 1 & 4 &  &  & X\\
PAYEMS & 1 & 4 & X & X & X\\
USGOOD & 1 & 4 &  &  & X\\
\addlinespace
CES1021000001 & 1 & 4 &  &  & X\\
USCONS & 1 & 4 &  &  & X\\
MANEMP & 1 & 4 &  &  & X\\
DMANEMP & 1 & 4 &  &  & X\\
NDMANEMP & 1 & 4 &  &  & X\\
\addlinespace
SRVPRD & 1 & 4 &  &  & X\\
USTPU & 1 & 4 &  &  & X\\
USWTRADE & 1 & 4 &  &  & X\\
USTRADE & 1 & 4 &  &  & X\\
USFIRE & 1 & 4 &  &  & X\\
\addlinespace
USGOVT & 1 & 4 &  &  & X\\
CES0600000007 & 1 & 4 & X & X & X\\
AWOTMAN & 1 & 4 &  &  & X\\
AWHMAN & 1 & 4 &  & X & X\\
CES0600000008 & 1 & 4 & X & X & X\\
\addlinespace
CES2000000008 & 1 & 4 &  &  & X\\
CES3000000008 & 1 & 4 &  &  & X\\
HOUST & 1 & 16 & X & X & X\\
HOUSTNE & 1 & 16 &  &  & X\\
HOUSTMW & 1 & 16 &  &  & X\\
\addlinespace
HOUSTS & 1 & 16 &  &  & X\\
HOUSTW & 1 & 16 &  &  & X\\
DPCERA3M086SBEA & 1 & 23 & X & X & X\\
M1SL & 1 & 10 &  & X & X\\
M2REAL & 1 & 15 &  &  & X\\
\addlinespace
AMBSL & 1 & 11 &  &  & X\\
TOTRESNS & 1 & 10 &  & X & X\\
NONBORRES & 1 & 10 &  &  & X\\
BUSLOANS & 1 & 11 &  & X & X\\
REALLN & 1 & 11 &  &  & X\\
\addlinespace
NONREVSL & 2 & 8 &  &  & X\\
MZMSL & 1 & 11 &  &  & X\\
DTCOLNVHFNM & 2 & 29 &  &  & X\\
DTCTHFNM & 2 & 29 &  &  & X\\
INVEST & 1 & 11 &  &  & X\\
\addlinespace
FEDFUNDS & 1 & 1 & X & X & X\\
TB3MS & 1 & 1 &  &  & X\\
TB6MS & 1 & 1 &  &  & X\\
GS1 & 1 & 1 &  &  & X\\
GS5 & 1 & 1 &  &  & X\\
\addlinespace
GS10 & 1 & 1 &  & X & X\\
AAA & 1 & 1 &  &  & X\\
BAA & 1 & 7 &  &  & X\\
TB3SMFFM & 1 & 1 &  &  & X\\
TB6SMFFM & 1 & 1 &  &  & X\\
\addlinespace
T1YFFM & 1 & 1 & X & X & X\\
T5YFFM & 1 & 1 &  &  & X\\
T10YFFM & 1 & 1 & X & X & X\\
AAAFFM & 1 & 1 &  &  & X\\
BAAFFM & 1 & 1 & X & X & X\\
\addlinespace
PPIFGS & 1 & 11 &  & X & \\
PPICMM & 1 & 11 & X & X & X\\
CPIAUCSL & 1 & 15 &  & X & X\\
CPIAPPSL & 1 & 15 &  &  & X\\
CPITRNSL & 1 & 15 &  &  & X\\
\addlinespace
CPIMEDSL & 1 & 15 &  &  & X\\
CUSR0000SAC & 1 & 15 &  &  & X\\
CUSR0000SAS & 1 & 15 &  &  & X\\
CPIULFSL & 1 & 15 &  &  & X\\
CUSR0000SA0L5 & 1 & 15 &  &  & X\\
\addlinespace
PCEPI & 1 & 23 & X & X & X\\
DDURRG3M086SBEA & 1 & 23 &  &  & X\\
DNDGRG3M086SBEA & 1 & 23 &  &  & X\\
DSERRG3M086SBEA & 1 & 23 &  &  & X\\
IPB51222S & 1 & 16 &  &  & X\\
\addlinespace
UEMP15T26 & 1 & 4 &  &  & X\\
AMDMUOx & 1 & 23 &  &  & X\\
BUSINVx & 2 & 15 &  &  & X\\
ISRATIOx & 2 & 15 &  &  & X\\
EXSZUSx & 1 & 1 &  &  & X\\
\addlinespace
EXJPUSx & 1 & 1 &  &  & X\\
EXUSUKx & 1 & 1 & X & X & X\\
EXCAUSx & 1 & 1 &  &  & X\\
RETAILx & 1 & 15 &  &  & X\\
S.P.500 & 1 & 1 & X & X & X\\
\addlinespace
S.P..indust & 1 & 1 &  &  & X\\
S.P.div.yield & 1 & 1 &  &  & X\\
S.P.PE.ratio & 1 & 1 &  &  & X\\
COMPAPFFx & 1 & 1 &  &  & X\\
HWI & 1 & 4 &  & X & X\\
\addlinespace
HWIURATIO & 1 & 4 &  &  & X\\
AMDMNOx & 1 & 15 &  &  & X\\
CLAIMSx & 1 & 1 &  &  & X\\
CONSPI & 1 & 1 &  & X & X\\
CP3Mx & 1 & 1 &  &  & X\\
\addlinespace
M2SL & 1 & 15 &  & X & X\\
WPSFD49207 & 1 & 15 & X &  & X\\
WPSFD49502 & 1 & 15 &  &  & X\\
WPSID61 & 1 & 15 &  &  & X\\
WPSID62 & 1 & 15 &  &  & X\\
\addlinespace
CUSR0000SAD & 1 & 15 &  &  & X\\
CUSR0000SA0L2 & 1 & 15 &  &  & X\\
OILPRICEx & 1 & 15 &  &  & X\\
CMRMTSPLx & 2 & 29 & X & X & X\\ 
\end{longtable}
}

\section{Prior and Posterior Distributions}
\label{app:prior}
\subsection{Prior Distributions}
\paragraph{Priors for the factor stochastic volatility model}
We use relatively loose priors for the parameters in the factor stochastic volatility model. The prior for the means of the idiosyncratic volatilities is $\mu\sim N(0, 10)$. To ensure that the volatility processes are stationary, we let $\frac{\phi+1}{2}\sim Beta(10, 3)$. The prior for the error variances of the volatility processes is $\sigma^2\sim\chi^2(1)$, and a standard normal prior is used for the factor loadings, i.e. $\Lambda_{ij}\sim N(0,1)$.
\paragraph{Minnesota-style prior for the regression parameters}
We set the prior mean for all regression parameters $\Pi_l^{(i, j)}$ to zero. The prior variance is set according to
\begin{align}
\sqrt{V\left(\Pi_l^{(i, j)}\right)}=\begin{cases}
\frac{\lambda_1}{l^{\lambda_3}}, &\quad \text{if } i = j\\
\frac{\lambda_1\lambda_2}{l^{\lambda_3}}\frac{s_i}{s_j}, &\quad \text{otherwise.}
\end{cases}
\end{align}
The overall tightness is set to $\lambda_1=0.2$ for CCM-20 and Koop-34, whereas CCM-119 uses $\lambda_1=0.1$. The cross-variable tightness and lag decay are set to $\lambda_2=0.5$ and $\lambda_3=2$ for all three models.

\subsection{Posterior Distributions}
We use MCMC to sample from the posterior. The target distribution is
\begin{align}
p(\phi, \mu, \sigma, \Lambda_f, f, \Pi, x, \Omega|y),
\end{align}
where $\phi, \mu, \sigma$ are the mean, AR(1) and standard deviation parameters from the univariate log-volatility regressions, $\Lambda_f$ are the factor loadings, $f$ are the factors, $\Pi$ the VAR regression parameters, $x$ the latent data and $\Omega$ the latent volatilities. To sample from the posterior distribution of the latent volatilities, we use the \cite{Kim1998} mixture representation, employing the 10-state mixture proposed by \cite{Omori2007}. Doing so introduces the mixture indicators $s$ to the target posterior, thereby obtaining
\begin{align}
p(\phi, \mu, \sigma, \Lambda_f, f, \Pi, x, s, \Omega|y).
\end{align}

In order to sample from the correct posterior, we sample the mixture indicators immediately before the latent volatilities $\Omega$, as pointed out by \cite{DelNegro2015}. The full MCMC procedure consists of the following steps:
\begin{align}
p(\phi, \mu, \sigma|\Omega)\\
p(\Lambda_f|f, x, \Omega)\\
p(f|\Lambda_f, x, \Omega)\\
p(\Pi|\Lambda_f, f, x, \Omega)\\
p(x|\Pi, \Lambda_f, f, \Omega, y)\\
p(s|\Pi, \Lambda_f, f, x, \Omega)\\
p(\Omega|s, x, \phi, \mu, \sigma)
\end{align}
The steps relating to the factor stochastic volatility model are sampled using the deep interweaving strategy developed by \cite{Kastner2016}. Our MCMC procedure is fully implemented in C++ via Rcpp \citep{Eddelbuettel2011}, where we employ an implementation of the factor stochastic volatility part that is adapted from the \texttt{R} package \texttt{factorstochvol} \citep{Kastner2018}.
 
The regression parameters $\Pi$ are sampled on an equation-by-equation (corresponding to row-by-row) basis based on \eqref{eq:xtilde}--\eqref{eq:phidecomp}. Further speed improvements can be obtained by exploiting the fact that the (conditionally normal) posterior has moments with a certain structure. Let $D_i$ be a diagonal matrix with its diagonal elements being equal to the prior variances of $\Pi^{(i,\cdot)}$, i.e. the regression parameters in equation $i$. The conditional posterior is $\Pi^{(i, \cdot)}|\ddot{X}_i,\ddot{x}_i\sim N(m_i, V_i)$ as discussed in the main text, where
\begin{align}
V_i&=(\ddot{X}_i'\ddot{X}_i+D_i^{-1})^{-1}\\
m_i&=V_i\ddot{X}_i'\ddot{x}_i.
\end{align}
We employ the algorithm by \cite{Bhattacharya2016} when $np>T$ and use the procedure presented by \cite{Rue2001} when $T<np$. For completeness, the algorithms are outlined below.\vspace{0.2em}

\noindent\begin{minipage}[t]{0.45\textwidth}
\textbf{\cite{Rue2001} sampler}
\begin{enumerate}[wide, labelwidth=!, labelindent=0pt]
\item Compute $V_i^{-1}=LL'$
\item Solve $Lv=\ddot{X}_i'\ddot{x}_i$ and $L'\mu=v$
\item Sample $z\sim N(0, I_{np+1})$
\item Solve $L'y=z$
\item Set $\Pi^{(i, \cdot)}=\mu + y$
  \end{enumerate}
\end{minipage}
\begin{minipage}[t]{0.5\textwidth}
\textbf{\cite{Bhattacharya2016} sampler}
\begin{enumerate}[leftmargin=*]
\item Sample $u\sim N(0, D_i)$, \\$\delta \sim N(0, I_T)$
\item Set $v=\ddot{X} u+\delta$
\item Solve $(\ddot{X}D_i\ddot{X}'+I_T)w=(\ddot{x}_i-v)$
\item Set $\Pi^{(i, \cdot)}=u+D_i\ddot{X}'w$.
  \end{enumerate}\vspace{3em}
\end{minipage}

The \cite{Rue2001} sampler exploits the structure of the posterior and uses forward and backward substitution for triangular matrices in order to improve over a vanilla implementation where $V_i$ and $m_i$ are computed explicitly. The complexity is $O(n^3p^3)$. The \cite{Bhattacharya2016} sampler instead makes use of the fact that in many instances $np>T$, so inversion of $T\times T$ matrices may be cheaper. As noted in the main text, the complexity of the algorithm is $O(T^2np)$.

\end{document}